\documentstyle[prd,aps,floats]{revtex}
\input epsf

\newcommand{\be}{\begin{equation}}
\newcommand{\ee}{\end{equation}}
\newcommand{\bea}{\begin{eqnarray}}
\newcommand{\eea}{\end{eqnarray}}

\begin{document}
\draft
\twocolumn[\hsize\textwidth\columnwidth\hsize\csname
@twocolumnfalse\endcsname
\preprint{PU-RCG-97/15,~SUSX-TH-97/16,~hep-th/9708154}
\title{Cosmology of the type IIB superstring}
\author{E. J. Copeland$^1$, James E. Lidsey$^2$ and David Wands$^3$}
\address{$^1$Centre for Theoretical Physics, University of Sussex, 
Brighton, BN1 9QH,\ U.K. \\
$^2$Astronomy Unit, School of Mathematical Sciences, Queen Mary and Westfield, 
Mile End Road, London, E1 4NS,\ U.K. \\
$^3$School of Computer Science and Mathematics, 
University of Portsmouth, Portsmouth PO1 2EG,\ U.K.}
\date{\today}
\maketitle
\begin{abstract}
The continuous and discrete symmetries of a dimensionally reduced type
IIB superstring action are employed to generate four--dimensional
cosmological solutions with non--trivial Neveu--Schwarz/Neveu--Schwarz
and Ramond--Ramond form--fields from the
dilaton--moduli--vacuum solutions.

\end{abstract}
\pacs{PACS numbers: 98.80.Cq, 11.25.-w, 04.50.+h \hspace*{1mm} 
PU-RCG-97/15,~SUSX-TH-97/16,~hep-th/9708154}

 \vskip2pc]

Superstring theory is the most promising candidate for a unified
quantum theory of the fundamental interactions including
gravity~\cite{gsw}. The realization that the five different string
theories are related non--perturbatively by duality symmetries has led
to renewed interest in the type IIB
theory~\cite{reviewduality,hulltownsend}.  The Ramond--Ramond (RR)
sector of this theory is particularly relevant in view of the recent
advances that have been made in understanding the relationship between
D--branes of open string theory and RR gauge fields~\cite{D}.  The
main emphasis to date has been on black hole and membrane
solutions~\cite{duff}. However,
cosmological models might provide one of the few observational tests
of the theory and the early universe represents a natural environment
in which concepts such as string duality may be quantitatively
studied, at least within the context of the low energy supergravity
action.

In this paper we will present four--dimensional (4--D) cosmological
solutions of the type IIB theory with time--dependent
Neveu--Schwarz/Neveu--Schwarz (NS--NS) and RR fields by exploiting the
SL(2,R) symmetries of the theory~\cite{hulltownsend,2bsym}.
Solutions with a single NS--NS or RR form--field can be directly
integrated to yield simple analytic
expressions~\cite{CLW,luovwa,lmp,CLW97}. Solutions with more than one
form--field have also been found where the system reduces to an
integrable Toda model~\cite{luovwa,lmp,kaloper}. Here we include the
interactions between the NS--NS and RR form--fields on the external
space, which in general leads to a more complicated system, but one
which is still integrable due to the symmetries between the
fields. Our results generalise solutions previously obtained by a
single SL(2,R) transformation acting on the NS--NS sector
solutions~\cite{ps,fen}.

The low energy limit of the type IIB superstring is ten--dimensional
(10--D) $N=2$ chiral supergravity~\cite{sh}. 
The NS--NS fields of the bosonic sector of this theory 
are the dilaton, $\Phi$, the 10--D metric, $g_{MN}$,
and the antisymmetric two-form potential, $B^{(1)}_{MN}$. 
The RR sector contains a scalar
axion field, $\chi$, a two-form potential, $B^{(2)}_{MN}$, and a
four-form potential $D_{MNPQ}$. 
The equation of motion for the latter field cannot be derived from a
covariant action~\cite{cov} but we can consistently set it to zero. 
The effective action for the remaining fields is given by~\cite{hull1}
\bea
\label{IIB}
S &=& \int d^{10} x \sqrt{-g_{10}} \left\{ e^{-\Phi} 
\left[ R_{10} +\left( \nabla \Phi \right)^2 
-\frac{1}{12} \left( H^{(1)} \right)^2  \right] \right. \nonumber \\
&& \left. -\frac{1}{2} \left( \nabla \chi \right)^2 -\frac{1}{12} 
\left(  H^{(1)} \chi +H^{(2)} \right)^2 \right\}  ,
\eea
where $R_{10}$ is the Ricci curvature scalar, 
$g_{10} \equiv {\rm det} |g_{MN}|$ and
$H^{(i)}_{MNP} 
\equiv \partial_{[M} B^{(i)}_{NP]}$ are the field strengths of 
the two--form potentials $B^{(i)}_{MN}$.

To investigate 
4--D cosmological solutions we compactify the 10--D
spacetime on an isotropic six-torus
\be
\label{ansatz}
ds_{10}^2 =g_{\mu\nu} (x) dx^{\mu} dx^{\nu} 
+e^{y(x)/\sqrt{3}} \delta_{ab} dX^a dX^b  ,
\ee
where $y(x)$ describes the volume of the internal space and is the only
modulus field considered. We neglect moduli fields
arising from the compactification of the form--fields on the internal
dimensions \cite{toroidal,mah,roy}. The symmetries of the reduced theory
become manifest in the conformally related 4--D Einstein frame
\be
\tilde{g}_{\mu\nu} = e^{-\varphi} g_{\mu\nu}   ,
\ee
where $\varphi \equiv \Phi - \sqrt{3}y$ is the 4--D dilaton. 
In four dimensions, the three-form
field strengths are dual to the gradients of pseudo-scalar axion fields, 
$\sigma_i$ \cite{clw}: 
\begin{eqnarray}
\label{defH1}
\tilde{H}^{(1) \, \mu\nu\lambda} &=& \tilde\epsilon^{\mu\nu\lambda\kappa}
 e^{2\varphi} \left( \tilde{\nabla}_\kappa\sigma_1 - 
\chi \tilde{\nabla}_\kappa\sigma_2
 \right) \\
\label{defH2}
\tilde{H}^{(2) \, \mu\nu\lambda} &=& \tilde\epsilon^{\mu\nu\lambda\kappa}
  \left[ e^{\varphi-\sqrt{3}y}\tilde{\nabla}_\kappa\sigma_2
 \right. \nonumber \\
&& \left. - \chi e^{2\varphi}
 \left( \tilde{\nabla}_\kappa\sigma_1 - \chi
\tilde{\nabla}_\kappa\sigma_2 \right)
 \right]   .
\end{eqnarray}
In this dual formulation the equations of motion for the fields 
follow from an effective action \cite{clw}:
\bea
\label{solitonicaction}
S&=&\int d^4 x \sqrt{-\tilde{g}} \left[ \tilde{R} -\frac{1}{2} 
\left( \tilde{\nabla} \varphi \right)^2 - {1\over2} \left( 
\tilde{\nabla} y \right)^2  
 \right. \nonumber \\
&& \left. 
-\frac{1}{2} e^{\sqrt{3}y +\varphi} \left( \tilde{\nabla} \chi \right)^2 
-\frac{1}{2} e^{-\sqrt{3}y +\varphi} \left( \tilde{\nabla} \sigma_2 \right)^2
 \right. \nonumber \\
&& \left. 
-\frac{1}{2} 
e^{2\varphi} \left( \tilde{\nabla} \sigma_1 -\chi 
\tilde{\nabla} \sigma_2 \right)^2 
\right]   .
\eea

The action in Eq.~(\ref{solitonicaction}) 
is invariant under the global SL(2,R) transformation~\cite{mah,roy}
\be
\label{sl2rchi}
\bar{M} = \Sigma M \Sigma^T, \ \ \bar{\tilde{g}}_{\mu\nu} = 
\tilde{g}_{\mu\nu}, \ \ \bar{\sigma} = \left( \Sigma^T \right)^{-1} \sigma, 
\ \ \bar{v} =v ,
\ee
where 
\bea
\label{ufield}
{1\over2}\Phi \equiv u \equiv \frac{1}{2} \varphi + {\sqrt{3}\over2} y, 
\hspace{1cm}
v \equiv \frac{\sqrt{3}}{2} \varphi - {1\over2} y \\
\label{matrixM}
M \equiv \left( \begin{array}{cc} e^u &  \chi e^u \\ 
\chi e^u & e^{-u} + \chi^2 e^u \end{array} \right) , \hspace{1cm}
\sigma \equiv \left( \begin{array}{c} -\sigma_1 \\ 
  \sigma_2 \end{array} \right) 
\eea
and 
\be
\label{sigma}
\Sigma \equiv \left( \begin{array}{cc} D & C \\ 
B  & A \end{array} \right) , \qquad AD-BC =1   .
\ee

The assumption that all fields are independent of the internal  coordinates
implies that the  action (\ref{solitonicaction}) 
also exhibits a `T--duality'
\bea
\label{td}
\bar{y} =-y , \qquad \bar{\sigma}_1 =-\sigma_1+\chi \sigma_2 \nonumber \\
\bar{\chi} = \sigma_2 , \qquad \bar{\sigma}_2 = \chi 
\eea
that inverts the volume of the internal 
space and leaves the dilaton and 4--D Einstein frame metric 
invariant \cite{clw}. 

This results in a second SL(2,R) symmetry 
that may be viewed as a mirror image
of that given in Eq.~(\ref{sl2rchi}) \cite{clw}. If we define
\bea
\label{wfield}
w \equiv \frac{1}{2} \varphi - {\sqrt{3}\over2} y, \hspace{1.5cm}
x \equiv \frac{\sqrt{3}}{2} \varphi + {1\over2} y \\
\label{matrixP}
P \equiv \left( \begin{array}{cc} e^w &  \sigma_2 e^w \\ 
\sigma_2 e^w & e^{-w} + \sigma_2^2 e^w \end{array} \right) \hspace{.5cm} 
\rho \equiv \left( \begin{array}{c} \sigma_1-\chi\sigma_2 \\ 
  \chi \end{array} \right)    ,
\eea
the action is invariant under the SL(2,R) transformation
\be
\label{sl2rsigma2}
\bar{P} = \Sigma P \Sigma^T, \ \ \bar{\tilde{g}}_{\mu\nu} = 
\tilde{g}_{\mu\nu} , \ \ \bar{\rho} = \left( \Sigma^T \right)^{-1} \rho, 
\ \ \bar{x} =x .
\ee
The transformation (\ref{sl2rsigma2}) is formally equivalent 
to the sequence of transformations given by 
Eq. (\ref{td}), followed by Eq. (\ref{sl2rchi}), followed again 
by Eq. (\ref{td}). 

It should be emphasized that 
neither of the symmetries (\ref{sl2rchi}) or (\ref{sl2rsigma2}) 
coincide with the SL(2,R) symmetry of the NS--NS sector alone, 
which mixes the 4--D dilaton and the
NS-NS axion \cite{NSdual}. This symmetry has 
been employed previously to derive 4--D cosmological solutions
with a non--trivial NS--NS form-field \cite{CLW,CEW97}, 
but is broken due to the interaction of the RR fields. 

We consider Friedmann--Robertson--Walker (FRW)
cosmologies with the homogeneous and isotropic 4--D line element
\be
ds^2 = a^2(\eta) \left( -d\eta^2 + d\Omega_\kappa^2 \right)    ,
\ee
where $a$ is the scale factor in the string frame and
$d\Omega_\kappa^2$ is the line element on the 3--space with constant
curvature $\kappa$. The field equations are given by
\begin{eqnarray}
\label{field1}
\varphi'' + 2{\tilde{a}'\over \tilde{a}} \varphi' =
  {1\over2} e^{\sqrt{3}y+\varphi} \chi'^2
  + {1\over2} e^{-\sqrt{3}y+\varphi} \sigma_2'^2
\nonumber \\
+ e^{2\varphi} (\sigma_1'-\chi\sigma_2')^2 \\
\label{field2}
y'' + 2{\tilde{a}'\over \tilde{a}} y' =
  {\sqrt{3}\over2} e^{\sqrt{3}y+\varphi} \chi'^2
  - {\sqrt{3}\over2} e^{-\sqrt{3}y+\varphi} \sigma_2'^2 \\
\label{field3}
\chi'' + \left( 2{\tilde{a}'\over \tilde{a}} 
+\sqrt{3}y' + \varphi' \right) \chi' =
\nonumber \\
- e^{-\sqrt{3}y+\varphi} \sigma_2' (\sigma_1'-\chi\sigma_2') \\ 
\label{field4}
\sigma_2'' + \left( 2{\tilde{a}'\over \tilde{a}} 
-\sqrt{3}y' + \varphi' \right) \sigma_2'
 = e^{\sqrt{3}y+\varphi} \chi' (\sigma_1'-\chi\sigma_2') \\ 
\label{field5}
(\sigma_1'-\chi\sigma_2')' + 2\left( {\tilde{a}'\over 
\tilde{a}} + \varphi' \right)
  (\sigma_1'-\chi\sigma_2') = 0
\end{eqnarray}
together with the Friedmann constraint
\bea
\label{friedmann}
12 \left[ \left({\tilde{a}'\over \tilde{a}}\right)^2 + 
\kappa  \right]
 = \varphi'^2 + y'^2 + e^{\sqrt{3}y+\varphi}\chi'^2 \nonumber \\
+ e^{-\sqrt{3}y+\varphi}\sigma_2'^2 + e^{2\varphi}
 (\sigma_1'-\chi\sigma_2')^2    ,
\eea
where $\tilde{a} \equiv a e^{-\varphi /2}$ and a prime denotes $d /d\eta$. 

\begin{figure}
\begin{center}
\leavevmode\epsfysize=5.5cm \epsfbox{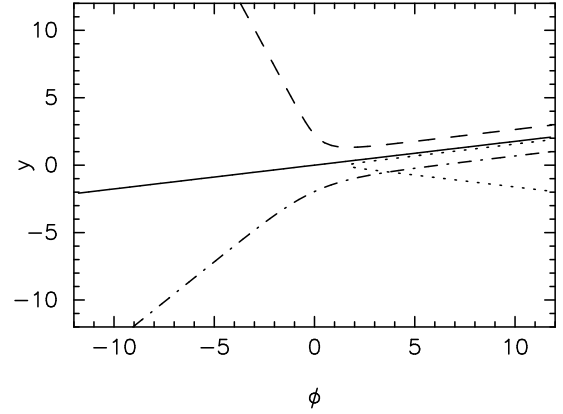}\\ 
\end{center}
\caption[Single form-field]
{Trajectories in $(\varphi,y)$ field--space for the 
dilaton--moduli--vacuum solution (solid line) with $\xi_*=\pi/9$. The
dashed, dot--dashed and dotted lines represent the three 
single form--field solution with $\psi=\chi$, $\sigma_2$ and
$\sigma_1$, respectively, obtained by the appropriate SL(2,R)
transformation of the dilaton--moduli--vacuum solution.}
\label{fig1}
\end{figure}

The general FRW dilaton--moduli--vacuum solution (with vanishing RR
fields and NS-NS three--form field strength) is given by~\cite{CLW}
\bea
\label{dilatonvacuum1}
a &=& a_* \sqrt{\tau^{1+\sqrt{3}\cos\xi_*} \over 1+\kappa\tau^2} \, \\
\label{dilatonvacuum2}
e^{\varphi} &=&e^{\varphi_*} \tau^{\sqrt{3}\cos\xi_*} \\
\label{dilatonvacuum3}
e^{y} &=&e^{y_*} \tau^{\sqrt{3}\sin\xi_*}   ,
\eea
where a subscript $*$ denotes arbitrary constants of integration
and $\tau$ is given in terms of the conformal time by
\begin{equation}
\tau \equiv \left\{
\begin{array}{ll}
\kappa^{-1/2} |\tan (\kappa^{1/2}\eta)| & {\rm for}\ \kappa>0 \\
|\eta| & {\rm for}\ \kappa=0 \\
|\kappa|^{-1/2} |\tanh (|\kappa|^{1/2}\eta)| & {\rm for}\ \kappa<0
\end{array}
\right. \ .
\end{equation}
The dilaton--moduli--vacuum solutions correspond to trajectories which
are straight lines in $(\varphi,y)$ field--space (see
Fig.~\ref{fig1}). Thus, vacuum trajectories which come in at an angle
$\xi_1$ to the $\varphi$ axis go out to infinity at an angle
$\xi_2=\xi_1+\pi$.
Note that the time coordinate $\tau$ diverges at early and late time in
models with $\kappa\geq0$, but in models with negative spatial curvature we
have $\tau\to|\kappa|^{-1/2}$ as $\eta\to\pm\infty$.

The scale factor in the Einstein frame is given by
\be
\label{Einsteina}
\tilde{a} = \tilde{a}_* \sqrt{\tau \over 1+\kappa\tau^2} \ .
\ee 
This follows directly from the fact that homogeneous massless scalar
fields have the same equation of state as a maximally stiff
fluid~\cite{MW95}.  This remains true when the form--fields are
included, despite their couplings to the dilaton and
moduli~\cite{CLW,CLW97}, since the 4--D Einstein frame metric is
invariant under the SL(2,R) transformations in Eqs.~(\ref{sl2rchi})
and~(\ref{sl2rsigma2}). Thus, the Einstein frame scale factor always
has the simple evolution described by Eq.~(\ref{Einsteina}) and this
is singular as $\eta\to0$ 
(and $\kappa^{1/2}\eta\to\pm\pi/2$ for $\kappa>0$).

The general FRW solution containing a single excited RR form
field~\cite{CLW,luovwa,lmp,CLW97} can be generated by applying the SL(2,R)
transformations in Eq.~(\ref{sl2rchi}) or Eq.~(\ref{sl2rsigma2}) to
the dilaton--moduli--vacuum solutions in
Eqs.~(\ref{dilatonvacuum1}--\ref{dilatonvacuum3}).  We obtain
\bea
\label{GENERALA}
a^{2n} &=& {a_*^{2n}\over2}
 \left[ { (\tau/\tau_*)^{n(1+\sqrt{3}\cos\xi_1)}  
+ (\tau/\tau_*)^{n(1-\sqrt{3}\cos\xi_2)} \over (1+\kappa\tau^2)^n}
 \right] \\
\label{GENERALPHI}
e^{\varphi} &=& {e^{\varphi_*} \over 2^{1/n}}
\left[ (\tau/\tau_*)^{n\sqrt{3}\cos\xi_1} 
+ (\tau/\tau_*)^{-n\sqrt{3}\cos\xi_2}
 \right]^{1/n} \\
\label{GENERALY}
e^{y} &=& {e^{y_*}\over2^{1/m}} \left[ (\tau/\tau_*)^{m\sqrt{3}\sin\xi_1}  
+ (\tau/\tau_*)^{-m\sqrt{3}\sin\xi_2} 
\right]^{1/m} \\
\label{GENERALFORM}
\psi &=&\psi_{*} + K^{-1} \left[
\frac{ (\tau/\tau_*)^{n\sqrt{3}\cos\xi_1}
 - (\tau/\tau_*)^{-n\sqrt{3}\cos\xi_2}}
 {(\tau/\tau_*)^{n\sqrt{3}\cos\xi_1}
 + (\tau/\tau_*)^{-n\sqrt{3}\cos\xi_2} } \right]   ,
\eea
where $K=\pm e^{(\varphi_*/n)+(y_*/m)}$ and the field $\psi$ 
represents the field $\chi$ or $\sigma_2$
depending upon which of these fields is excited.
These solutions interpolate between two asymptotic regimes
where the form--fields vanish and the trajectories 
in $(\varphi ,y)$ space become straight lines~\cite{CLW}.  
If the asymptotic trajectory comes in at an initial angle $\xi_1$ to the
$\varphi$ axis, it leaves at an angle $\xi_2$. 
The values of the parameters $n$, $m$, $\xi_1$ and $\xi_2$ for
different choices of form--field are given in Table~\ref{table1}. Note 
that for each form--field there is a characteristic angle $\theta$ such 
that $1/n = \cos \theta$, $1/m = \sin \theta$ and 
$\xi_2 = 2\theta - \xi_1$.    

\begin{table}
\caption[table]{Parameters in the cosmological solutions of 
Eqs.~(\ref{GENERALA}--\ref{GENERALY}) for different
choices of the $\psi$ field in Eq.~(\ref{GENERALFORM}).}
\begin{tabular}{|*{5}c|}
\hline
$~~~\psi$~~~ 	& ~~~$1/n$~~~ & ~~~$1/m$~~~ & ~~~$\xi_1$~~~ &
~~~$\xi_2$~~~ \\ 
\hline\hline
$\chi$		& $1/2$ & $\sqrt{3}/2$ & $\xi_*$ &
$(2\pi/3)-\xi_*$~~ \\
\hline 
$\sigma_2$	& $1/2$ & $-\sqrt{3}/2$ 	& $\xi_*$ & $-(2\pi/3)-\xi_*$ \\ 
\hline
$\sigma_1$ 	& $1$ & $0$ 		& $\xi_*$ & $-\xi_*$ \\ 
\hline
\end{tabular}
\label{table1}
\end{table}

The general solution with non-trivial $\chi$ and constant $\sigma_i$  
(i.e., vanishing three--form field strengths $H^{(i)}$)
is obtained by applying the SL(2,R) transformation (\ref{sl2rchi}).
The transformed fields $\bar{u}$ and $\bar\chi$ have the form
\bea
\label{baru}
e^{\bar{u}} &=& \left| 2 C (D+C\chi) \right| \cosh \left( u + \Delta
\right) \ ,\\
\bar\chi &=& \chi_* \pm {1 \over |2C(D+C\chi)|} \tanh \left( u + \Delta
\right) \ ,
\eea
where $e^\Delta\equiv|(D+C\chi)/C|$. The introduction of a
non--constant $\bar\chi$ field places a lower bound on $\bar{u}$, and
hence the 10--D dilaton field, $\Phi=2u$.  A typical solution with
$\chi'\neq0$ is shown in Fig.~\ref{fig1}.  
The RR field interpolates between two
asymptotic vacuum solutions where $\chi'\to0$. 
Trajectories that come in 
from infinity ($u\to\infty$) at an angle $\xi_1=\xi_*$, where
$-\pi/6\leq\xi_*\leq5\pi/6$, are then reflected in the line $u=u_*=
(\varphi_* + \sqrt{3}y_*)/2$ and emerge at an angle $\xi_2=
(2\pi/3)-\xi_*$. 

The mirror image of the $\chi'\neq0$ solution under the reflection
symmetry Eq.~(\ref{td}) is a solution with $\sigma_2' \neq 0$ and
$\chi'=0$. Because the T-duality,
Eq.~(\ref{td}), leaves the 4--D dilaton $\varphi$ as well as the 4--D
Einstein frame metric invariant, we find that the evolution of both
$\varphi$ and thus the original string frame metric is the same for a
single excited RR field regardless of whether it is $\chi$ or
$\sigma_2$ that is excited. 

This solution with $\sigma_2' \neq 0$, can also be generated by applying the
transformation (\ref{sl2rsigma2}) to the dilaton--moduli--vacuum
solutions in Eqs.~(\ref{dilatonvacuum1}--\ref{dilatonvacuum3}). It leaves 
$\chi'=0$  but leads to $\sigma_1' = \chi \sigma_2'$ and thus
$H^{(1)}=0$, but $H^{(2)}\neq0$. 
Note that $\sigma_1$ is only constant when $\chi=0$. There is 
a lower bound on the field $w$, and the generic behaviour of
$\varphi$ and $y$ for this solution is plotted in Fig.~\ref{fig1}.  
The RR form--field again leads to a solution that
interpolates between two asymptotic vacuum solutions, where
$\sigma_2'\to0$. Trajectories that come from 
infinity ($w\to\infty$) at an angle $\xi_1=\xi_*$, where 
$-5\pi/6\leq\xi_*\leq\pi/6$, are reflected in the line $w=w_*=
(\varphi_* - \sqrt{3}y_*)/2$, and emerge at an angle $\xi_2= -(2\pi/3)-\xi_*$. 

For completeness we note that Eqs.~(\ref{GENERALA}--\ref{GENERALFORM})
also represent a solution with $\psi=\sigma_1$ (see
Table~\ref{table1}) which is the general `dilaton--moduli--axion'
solution first presented in~\cite{CLW}. For $\chi=0$ this corresponds
to vanishing RR field strengths and an excited NS--NS three-form field
strength, $H^{(1)}$.  Note that for $\chi\neq0$ (but constant) this
corresponds to a particular solution with non-vanishing RR three-form
field strength $H^{(2)}=-\chi H^{(1)}$ [see Eqs.~(\ref{defH1})
and~(\ref{defH2})].  The scale factor in the Einstein frame is still
given by Eq.~(\ref{Einsteina}) and the modulus field $y$ is given by
the vacuum solution (\ref{dilatonvacuum3}).  The typical evolution of
the fields $\varphi$ and $y$ is shown in Fig.~\ref{fig1}.
The presence of a non-vanishing $\sigma_1'$ enforces a lower bound on
the value of the 4--D dilaton, $\varphi\geq\varphi_*$. Trajectories
that come from infinity ($\varphi\to\infty$) at an angle
$\xi_1=\xi_*$, where $-\pi/2\leq\xi_*\leq\pi/2$, are reflected in the
line $\varphi=\varphi_*$ back out at an angle $\xi_2=-\xi_*$. No
trajectories can reach $\varphi\to-\infty$ unless
$\sigma_1'=0$~\cite{CLW}.

We now consider solutions where two of the form--fields are non-vanishing 
but the third is zero. The field equations
(\ref{field1}--\ref{field5}) imply 
that the only consistent solution of this type arises when 
$H^{(1)}=0$. {}From Eq.~(\ref{defH1}), $\sigma_1' = \chi
\sigma_2'$ and this allows $\sigma_1$ to be eliminated. 
Eqs.~(\ref{field3}) and~(\ref{field4}) may be integrated directly 
to yield $\tilde{a}^2e^{\varphi+\sqrt{3}y}\chi'=L$ 
and $\tilde{a}^2e^{\varphi-\sqrt{3}y}\sigma_2'=J$,
where $J$ and $L$ are arbitrary constants. Defining 
a new time parameter $T\equiv \int^{\eta} d\eta'/ \tilde{a}^2
\propto \ln | \tau |$  
and new variables $q_{\pm} \equiv \varphi \pm(y/\sqrt{3})$
implies that the field
equations for the dilaton and moduli may be expressed as 
\bea
\label{fieldq1}
\ddot{q}_- =J^2 e^{q_+ -2q_-} \\
\label{fieldq2}
\ddot{q}_+ = L^2 e^{q_- -2q_+}
\eea
and the Friedmann constraint (\ref{friedmann}) gives
\be
\label{friedmannq}
{1\over8} \left( \dot{q}_+ + \dot{q}_- \right)^2
+ {3\over8} \left( \dot{q}_+ - \dot{q}_- \right)^2 + V
=\frac{3a_*^4}{2}  ,
\ee
where a dot denotes $d/dT$ and the potential 
\be
V \equiv  {1\over2}\left(J^2 e^{q_+ -2q_-} + L^2 e^{q_- -2q_+}\right)
\ee

Equations~(\ref{fieldq1}--\ref{friedmannq}) correspond to those of the
SU(3) Toda system~\cite{toda}. This  has recently been studied in
similar models by a number of authors~\cite{luovwa,lmp,kaloper}.
The general solution is of the form~\cite{kaloper}
\be
e^{q_-} = \sum_{i=1}^3 A_i e^{-\lambda_iT} \ , \quad
e^{q_+} = \sum_{i=1}^3 B_i e^{\lambda_iT} \ ,
\ee
where $\sum_i\lambda_i=0$, so that $\lambda_{\rm
min}<0$ and $\lambda_{\rm max}>0$. 
This gives the asymptotic solution for $\varphi$ and $y$ as
$T\to-\infty$:
\be
e^\varphi \sim e^{-(\lambda_{\rm max}-\lambda_{\rm min})T/2}
\ ,\quad 
e^y \sim e^{\sqrt{3}(\lambda_{\rm max}+\lambda_{\rm min})T/2}  ,
\ee
while as $T\to+\infty$ we have
\be
e^\varphi \sim e^{(\lambda_{\rm max}-\lambda_{\rm min})T/2}
\ ,\quad
e^y \sim e^{\sqrt{3}(\lambda_{\rm max}+\lambda_{\rm min})T/2}  .
\ee

As in the single form-field solutions discussed above, the asymptotic
solutions correspond to straight lines in the $(\varphi,y)$ plane.  We
see that trajectories that come from infinity ($\varphi\to\infty$) at
an angle $\xi_*$ are reflected back out at an angle $-\xi_*$. This is
exactly the qualitative behaviour of the NS--NS dilaton--moduli--axion
solution.  However, the range of allowed asymptotic trajectories is
more restricted than in the pure NS--NS case. The potential in the
constraint Eq.~(\ref{friedmannq}) is bounded from above and we
therefore require that $|y|\leq\varphi/\sqrt{3}$ asymptotically. Thus,
the asymptotic solutions are restricted to the range
$-\pi/6\leq\xi_*\leq\pi/6$, where $V\leq3a_*^4/2$.

The general FRW solutions to the type IIB string action presented in
Eq.~(\ref{solitonicaction}) can be generated by applying an appropriate
sequence of SL(2,R) transformations on the dilaton--moduli--vacuum
solution presented in Eqs.~(\ref{dilatonvacuum1}--\ref{dilatonvacuum3}).
The transformations are given by Eq. (\ref{sl2rchi}), followed by Eq.
(\ref{sl2rsigma2}), followed by Eq. (\ref{sl2rchi}). This includes the
above Toda system as a special case, where $H^{(1)}=0$. 
The procedure is readily extended to other situations including
anisotropic and inhomogeneous metrics.
One can also obtain solutions with all three form--fields
non--zero~\cite{ps,fen} by applying the SL(2,R) transformation given in
Eq.~(\ref{sl2rchi}) to the single form--field $\psi=\sigma_1$
solutions presented in
Eqs.~(\ref{GENERALA}--\ref{GENERALFORM}). However,  these solutions
correspond only to the particular case where $H^{(2)}\propto H^{(1)}$.

\begin{figure}
\begin{center}
\leavevmode\epsfysize=5.5cm \epsfbox{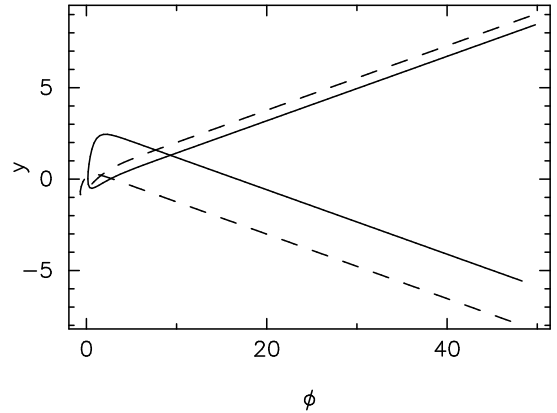}\\ 
\end{center}
\caption[Two and three form fields]
{Trajectories in $(\varphi,y)$ field--space for a typical
solution with two RR form--fields with $H^{(1)}=0$ (dashed line) and a
solution with all form--fields non--trivial (solid line). 
The two solutions are related by an SL(2,R) transformation.}
\label{fig2}
\end{figure}

\begin{figure}
\begin{center}
\leavevmode\epsfysize=5.5cm \epsfbox{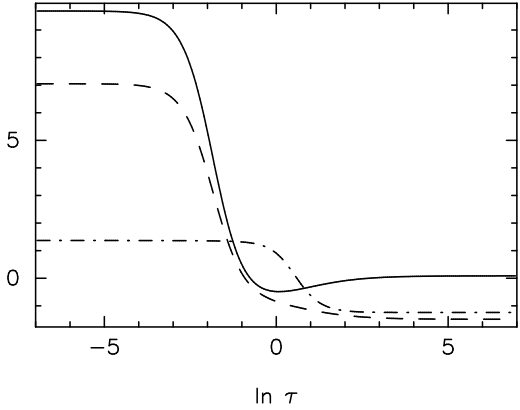}\\ 
\end{center}
\caption[Axion fields]
{The three axion fields $\sigma_1$ (solid line), $\sigma_2$
(dashed line) and $\chi$ (dot--dashed line) against $\ln\tau$ 
for the solution shown in Fig.~2 with all form--fields non--trivial.}
\label{fig3}
\end{figure}

The general solution exhibits a sequence of bounces between asymptotic
vacuum states.  A typical solution is shown in Figs.~\ref{fig2}
and~\ref{fig3}. The time--dependence of the fields $\chi$ and
$\sigma_2$ induces lower bounds on the variables $u$ and $w$,
respectively, as seen in the single form--field solutions.  
In the general solution this results in a
lower bound on $\varphi = u+w$.

The general type IIB solution contains a non-vanishing NS--NS
form--field, but can always be obtained from a Toda system with
$H^{(1)}=0$ by a single SL(2,R) transformation (\ref{sl2rchi}).  The
asymptotic behaviour of $\varphi$ and $y$ is invariant under this
transformation. This follows since  $u\to\infty$ asymptotically for all 
solutions in the Toda system\footnote{An exceptional case is when $u\to
u_*$ asymptotically, where $u_*$ is a  constant. In this case
$\bar{u}\to$constant, though not necessarily $u_*$, but the qualitative
behaviour is the same.}
and, from Eq.~(\ref{baru}), we obtain $\bar{u}\to u$ in the general
solution. We also have $\bar{v}=v$ and thus $\varphi$ and $y$ are
invariant in this limit. Thus, trajectories in $(\varphi,y)$ field-space
come in at an angle $\xi_*$ and leave at an angle $-\xi_*$, where
$-\pi/6\leq\xi_*\leq\pi/6$.

In conclusion, therefore, we have shown that from the dimensionally
reduced type IIB superstring action (\ref{solitonicaction}), the general
FRW cosmological solution with non-vanishing RR and NS--NS form--fields
interpolates between asymptotic dilaton--moduli--vacuum solutions, where
the form--fields vanish. These early-- and late--time limiting solutions
correspond to straight lines in the $(\varphi,y)$ plane at an angle
$\xi_*$ to the $\varphi$--axis and are related via a reflection symmetry
$\xi_*\to-\xi_*$.  This is strikingly similar to the pure NS--NS
solution\cite{CLW}. On the other hand, when the RR fields are
non--vanishing, the initial and final trajectories in $(\varphi ,y)$
space are restricted to the wedge $-\pi/6\leq\xi_*\leq\pi/6$, in
contrast to the NS--NS solutions, where they are bounded by
$-\pi/2\leq\xi_*\leq\pi/2$.

This places important restrictions on the range of 
cosmological solutions found. Firstly, the scale factor in the string
frame, $a$, is always bounded from below. The solutions are still
singular in this frame, however, and approach a curvature singularity
in the limit $\tau\to0$.  When the RR fields vanish, the axion field
$\sigma_1$ is dual to the NS--NS three-form field strength.  This
field is minimally coupled in the conformally related frame with scale
factor $e^{\varphi}a$. The metric in this frame is {\em
non--singular} when $|\xi_*|\leq\pi/6$~\cite{CEW97}. It is interesting
that this is precisely the bound placed on the allowed trajectories by
the RR fields. Moreover, the spectrum of quantum fluctuations induced
in the NS--NS axion field during an inflationary ``pre-Big Bang''
epoch~\cite{pbb} has a spectral index given by $n_s
= 4-2\sqrt{3}|\cos\xi_*|$~\cite{CLW97,CEW97}.  
The RR fields limit the allowed range to
$4-2\sqrt{3}\leq n_s\leq1$. Thus, RR fields in this model lead to the
NS--NS axion possessing a spectrum with $n_s\leq1$ and this might have
interesting consequences for the formation of large-scale structure in
our universe.

\end{document}